\documentclass[12pt]{iopart}
\usepackage{multirow}
\usepackage{iopams}
\usepackage{amsmath, amssymb, amstext}
\usepackage{cite}
\usepackage{caption, subcaption}
\usepackage{latexsym}
\usepackage{graphicx}
\usepackage{hyperref}
\usepackage{color}
\usepackage{enumerate}
\usepackage{tabularx}
\usepackage[normalem]{ulem}

\definecolor{bluc}{cmyk}{1,1,0,0.1}
\definecolor{rossoCP3}{cmyk}{0,.88,.77,.40}
\definecolor{rosso}{cmyk}{0,1,1,0.4}
\definecolor{rossos}{cmyk}{0,1,1,0.55}
\definecolor{rossoc}{cmyk}{0,1,1,0.2}
\definecolor{verdes}{cmyk}{0.92,0,0.59,0.4}
\hypersetup{colorlinks, bookmarksopen, bookmarksnumbered, citecolor=verdes, linkcolor=bluc, pdfstartview=FitH, urlcolor=rossos}

\usepackage{etoolbox}
\makeatletter
\newrobustcmd{\fixappendix}{%
  \patchcmd{\l@section}{1.5em}{7em}{}{}%
  \patchcmd{\l@subsection}{2.3em}{7em}{}{}%
}
\makeatother

\newcommand\rst{\bgroup\markoverwith{\textcolor{red}{\rule[0.5ex]{1pt}{1.4pt}}}\ULon}

\begin{document}
\title{Constraining modified gravity models through strong lensing cosmography}

\author{Mario H. Amante$^1$, Andrés Lizardo$^{1*}$, Javier Chagoya$^1$, C. Ortiz$^1$.}
\ead{$^*$andres.lizardo@fisica.uaz.edu.mx}
\address{Unidad Acad\'emica de F\'isica, Universidad Aut\'onoma de Zacatecas, C.P. 98060, Zacatecas, M\'exico.}
\date{\today}

\begin{abstract}
We analyze cosmography as a tool to constrain modified gravity theories. We take four distinct models and obtain their parameters in terms of the cosmographic parameters favored by observational data of strong gravitational lensing. We contrast with the values obtained by direct comparison between each model and the observational data. In general, we find consistency between the two approaches at 2$\sigma$ for all models considered in this work. Our study bridges the gap between theoretical predictions of modified gravity and empirical observations of strong gravitational lensing, providing a simple methodology to test the validity of these models.

\end{abstract}

\maketitle

\section{Introduction}

A revolution in cosmology began when observational data, such as Type Ia supernovae \cite{Riess:1998}, revealed that the Universe is currently undergoing a phase that appears to involve accelerated expansion of spacetime. Amongst the many efforts to explain this accelerated expansion, the most accepted proposal is the $\Lambda$CDM model, also known as the standard cosmological model, which considers that the constituents of the Universe, in addition to those dictated by the standard model of particle physics, are cold dark matter (CDM) and a cosmological constant ($\Lambda$) that is responsible for the present day expansion of the Universe. This model is consistent with cosmic microwave background (CMB) physics~\cite{Planck:2018vyg} and large scale structure observations \cite{Addison:2013haa, Nadathur:2020kvq}. However, recent measurements indicate some tension between different estimations of both the Hubble constant $H_0$ and the $S_8$ parameter, depending on whether these are based on the CMB or on measurements in the nearby Universe. These enduring and persistent tensions display disparities of up to $6 \sigma$ for $H_0$ and around 3$\sigma$ for the $S_8$ parameter when some dataset are considered (see \cite{Abdalla:2022yfr,DiValentino:2021izs,Vagnozzi:2023nrq,Hu:2023jqc,Perivolaropoulos:2021jda} for an extensive analysis). Furthermore, there is a long standing theoretical issue with the $\Lambda$CDM model. If $\Lambda$ is interpreted as vacuum energy, then Quantum Field Theory predicts a value for $\Lambda$ that deviates from the observational value by several orders of magnitude. These anomalies have been part of the motivation to formulate alternative explanations for the gravitational behavior of our Universe. These alternatives comprise an extensive range of possibilities, broadly divided into those that alter the energy-momentum tensor and those that modify the geometrical component of the Einstein equations; see, for example, \cite{Kobayashi:2011nu, Brax:2013ida, Joyce:2014kja, Li,Hu:2024qnx,Hu}. Apart from satisfying certain theoretical requirements, these alternative theories should be tested through comparison with observational data, which constantly improve in their precision. 

Rather than testing alternatives to GR and $\Lambda$CDM on a model-by-model basis, it is convenient to have a framework to analyze the dynamics of the Universe without any assumptions on the underlying model of gravity. This is the foundation of cosmography, which analyzes the Hubble factor in a Taylor expansion around the present time. The coefficients of the expansion -- so called cosmographic parameters --  are related to the derivatives of the scale factor at the present time. The goal is thus to derive the values of the cosmographic parameters from local observations. Several analyses on the usefulness of cosmography as a model independent test of gravity have been reported in the literature~(e.g. \cite{Capozziello:2008qc,Bamba:2012cp,Busti:2015xqa,Zhang_2017,TeppaPannia:2018ale,Lusso:2019akb,Camarena:2019moy,Mandal:2020buf}). A modification of the method that has attracted some attention is dubbed $y$-redshift,  {initially proposed in \cite{Catto_n_2007}}, which provides the theoretical advantage of mapping the redshift $z\in[0,\infty)$ to a parameter $y\in[0,1)$. In a similar spirit, a logarithmic expansion was presented in~\cite{Bargiacchi:2021fow}. These approaches have been used to estimate cosmographic parameters with different observational data sets~(e.g.~\cite{Lizardo:2020wxw, Magana:2017nfs, Fortunato:2023deh,Bargiacchi:2023rfd}).  

In this work, we revisit the question of whether cosmography is a viable approach to reconstruct alternative models of cosmology. {It is to be expected that not all models can be covered by cosmography as one considers larger values of $z$~\cite{OColgain:2021pyh}. Here we study this issue in detail using } an updated catalog of strong gravitational lenses (SLS, Strong Lensing Systems)~\cite{Amante:2019xao}  as a tool for constraining cosmological and  cosmographic parameters. {First, we obtain the best fit parameters for each model, then we perform some tests to analyse whether these models are compatible with cosmography, and finally we try to reconstruct the models starting from the best fit cosmographic parameters.}  Our methodology is based on a proposal for inferring cosmological parameters using elliptical galaxies as gravitational lenses~\cite{Grillo:2007iv}, which has been extended in the literature to include various lensing systems at different scales~\cite{Jullo:2010dq,Magana:2015wra,Magana:2017gfs,Caminha:2021iwo, Sonnenfeld:2021bhg, Moresco:2022phi, Wu:2022dgy, Diao:2022fre, Qi:2022kfg, Birrer:2022chj, Treu:2022aqp,Verdugo:2024dgs}. It is also worth mentioning that lensing observations hold promise as a valuable tool for analyzing the distribution of dark matter in galaxies and populations of galaxies~\cite{Chiba:2001wk, Hoekstra:2003pn, Massey:2010hh, Kaiser:1992ps}.

This work is organized as follows.  In Sec.~\ref{BCosm} we present the alternative cosmological models under consideration. 
In Sec.~\ref{sec:data}, we present the data and methodology to be used
to analyze the cosmological models using strong lensing observations. We also discuss the methodology to assess the viability of cosmography for reconstructing alternative cosmological models with strong lensing systems. In Sec.~\ref{sec:res}, we report the constraints for the free parameters of each {model}, and we also present the novel contrast with cosmography. In Sec.~\ref{sec:dis} we 
give our concluding remarks
and discuss some possibilities for future studies.

\section{Models of background cosmology} \label{BCosm}
In this section we present the cosmological models to be used in order to assess the validity of cosmography for constraining cosmological parameters using gravitational lenses. These models come from different theoretical approaches, including the $\omega$CDM parametrization, a holographic inspired model (hFRW)~\cite{Cai:2017asf}, a subset of degenerate higher order scalar tensor theories (DHOST)~\cite{Langlois:2015cwa,BenAchour:2016fzp}, and a model that considers coupling between gravity and the energy-momentum tensor ($f(R,T)$)~\cite{Harko:2011kv}\footnote{Not to be confused with theories of gravity that consider torsion, usually denoted $F(R,T)$.}.

\subsection{$\omega$CDM}
The $\omega$CDM model is an extension of $\Lambda$CDM
where the equation of state for dark energy is constant, but can be different from $\omega= -1$.
The equation of state should satisfy the condition $\omega < - \frac{1}{3}$ to produce an accelerated expansion of the Universe. The dimensionless Friedmann equation, $E(z)=H(z)/H_0$, for a universe consisting of baryonic matter, cold dark matter, and dark energy parametrized by $\omega$  can be written as
\begin{eqnarray}
E(z)_{\omega CDM}^2=\Omega_{0m}(1+z)^3+(1 -\Omega_{0m})(1+z)^{3(1+\omega)},\label{eq:ezw}
\end{eqnarray}
where $\Omega_{0m}$ {is}  the matter density parameter of baryonic and cold dark matter at $z=0$. Also, we are considering a flat universe, and the constraint $E(0) = 1$ has been imposed. 

In the next section, we use the normalized Hubble parameter $E(z)$ to obtain the angular diameter distance.
\subsection{hFRW}
The holographic FRW (hFRW) model consists of a FRW brane embedded in a flat bulk with an extra dimension relative to the brane. In this model, the stress-energy tensor of dark matter and dark energy is determined by the holographic stress-energy tensor on the brane hypersurface \cite{Cai:2017asf}. The dimensionless Friedmann equation can be written as 
\begin{eqnarray}
E(z)_{hFRW} & = & \left\{ \frac{\Omega_{0\lambda}}{2} + \Omega_{0B} \left( 1 + z \right)^3 \right. \nonumber \\
& &\left.  +\frac{\Omega_{0\lambda}}{2} \left[1 + \frac{4 \Omega_{0B}}{\Omega_{0\lambda}} \left( 1 + z \right)^3 + \frac{4 \Omega_{0I}}{\Omega_{0\lambda}} \left( 1 + z  \right)^4 \right]^{1/2}\right\}^{1/2}\, , \label{3hfrw}
\end{eqnarray}
where $\Omega_{0\lambda}$ is the dark energy density parameter, $\Omega_{0B}$ {is} the baryonic matter density parameter and $\Omega_{0I}$ is a density parameter related with an integration constant $I$ coming from the {metric} field equations in {the} hFRW scenario (see \cite{Cai:2017asf,Cai:2018ebs} for details). Imposing the flatness condition $E(0) =1 $, we obtain $\Omega_{0I} = \left( 1 - \Omega_{0\lambda} - 2 \Omega_{0B} + \Omega_{0B}^2\right)/\Omega_{0\lambda}$. Using this constraint to reduce the number of free parameters, equation \eqref{3hfrw} becomes
\begin{eqnarray}
E(z)_{hFRW} = \sqrt{\frac{\Omega_{0\lambda}}{2} + \Omega_{0B} \left(1 + z \right)^3 + Q^{\frac{1}{2}}}, \label{2hfrw}
\end{eqnarray}
where
\begin{equation}
Q= {\Omega_{0\lambda}^2}/{4} + \Omega_{0\lambda} \Omega_{0B} \left( 1 + z \right)^3 +  \left( 1 - \Omega_{0\lambda} - 2 \Omega_{0B} + \Omega_{0B}^2\right) \left( 1 + z \right)^4 .
\end{equation}
It is worth noting that in the case $\Omega_{0I} =0$ we recover the Friedmann equation of the self-accelerating branch of the Dvali-Gabadadze-Porrati model, which can also be obtained from entropic gravity (see \cite{Chagoya:2023hjw} for details).

\subsection{DHOST}
\begin{figure}
 \includegraphics[scale=0.75]{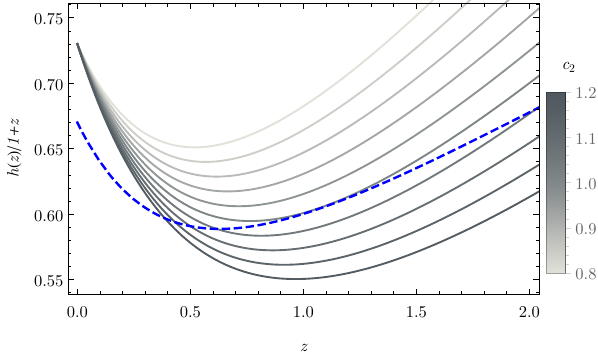}
\includegraphics[scale=0.75]{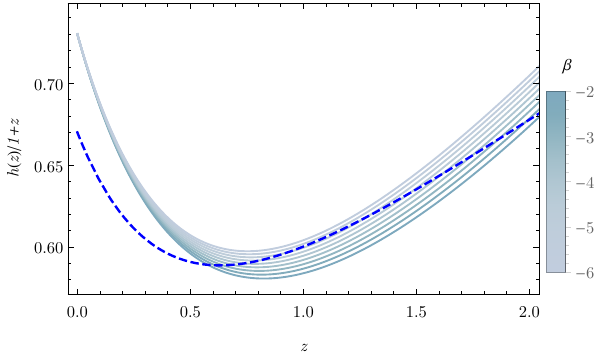} 
\caption{Numerical solutions for the reduced Hubble parameter as a function of redshift. In the left panel we fix $\beta =-5.3, c_3=5, c_4=1$ and $c_2\in[0.8,1.2]$, while in the right panel, we fix $c_2 = 1, c_3=5, c_4=1$ and $\beta\in[-6,-2]$.  {The dashed blue line shows $h(z)/1+z$ but in the $\Lambda$CDM scenario with $h_0=0.67$ and $\Omega_{0m}=0.315$.}\label{fig:hzdhost}}
\end{figure}

Degenerate Higher Order Scalar-tensor Theories (DHOST) are generalizations
of Horndeski gravity that allow for equations of motion of order higher than two, as long as a condition
to avoid propagation of Ostrogadski instabilities is satisfied.  We use a subsector
of this theory that is compatible
with luminal propagation of
gravitational waves, and has been shown to admit scaling solutions that mimic matter domination at early times and dark energy domination at late times. The general solution for the Hubble parameter in this sector is obtained numerically~\cite{Crisostomi:2017pjs} and depends on four\footnote{A fifth parameter, $\Lambda_3$, can be absorbed in a change of variables that makes the Hubble factor dimensionless. We account for this by normalizing the solution so that $h_0=0.74$, which is compatible with local observations \cite{Riess:2016jrr,Riess:2021jrx}.}  free parameters that appear in the free functions of the DHOST Lagrangians (see \ref{appendix:b} for details). We constructed a set of solutions for several values of two of these parameters, named $\beta$ -- which parametrises the difference between DHOST and beyond Horndeski, and $c_2$ -- which is associated with the canonical kinetic term of the scalar field, while the remaining two parameters are  {fixed as $c_3=5, c_4=1$ in order to explore a region of parameter space near the one used in~\cite{Crisostomi:2017pjs} }. Some examples of the profiles for $h(z)/1+z$ are presented in Fig.~\ref{fig:hzdhost}.

\subsection{$f(R,T)$ }
{The framework of $f(R,T)$ gravity is an extension of GR which allows for matter-curvature coupling in the Universe. The action is defined as follows: 
\begin{equation}
S = \int \left( \frac{f(R,T)}{16 \pi G} + S_{m}\right) \sqrt{-g} d^4x,
\end{equation}
where $S_m$ is the action for matter, $R$ the Ricci scalar and $T$ the trace of the energy-momentum tensor associated to $S_m$.  {$f(R,T)$ theories explore the consequences of matter-geometry couplings. In these theories, the energy-momentum tensor is not \textit{a priori} conserved, instead, one has an energy-balance equation that describes how the different matter components transfer energy between each other and also their possible interaction with the geometry. This property of $f(R,T)$ gravity can be used to model exotic fluids or quantum effects~\cite{Harko:2011kv}.}  An exact cosmological solution under the assumption $f(R,T) = \lambda(R + T)$ has been derived in \cite{Nagpal:2018mpv}, with $\lambda$ an arbitrary coupling constant of $f(R,T)$ gravity. The authors employ a parametrization\footnote{The choice of this parametrization arises from the fact that, in $f(R,T)$, two independent equations are obtained in terms of three unknowns. Therefore, the parameter $n$ constrained by the data is related to the physical parameters of the model.} of the Hubble parameter compatible with the existence of accelerating and decelerating phases in the evolution of the Universe. The corresponding dimensionless Friedmann equation has the form
\begin{equation}
E(z)= \frac{1}{2} \left( 1 + \left(1 + z  \right)^n \right),    \label{eq:ezfrt}
\end{equation}
where $n$ is parameter related to the physical parameters of the $f(R,T)$ scenario. }

\section{Data and Methodology} \label{sec:data}
Our methodology consists of two main steps: {we show that the models under consideration are compatible with cosmography, and then we try to reconstruct them from the model independent cosmographic parameters}. In the first step, we take the dimensionless Friedmann equation for the models described in the previous sections, we fit their parameters using {SLS}, we compute their theoretically related cosmographic parameters, and  we compare them with the actual, model independent, cosmographic parameters estimated from SLS. {In the second step, we obtain the parameters of each model from the model independent cosmographic parameters, and compare to their actual best fits to SLS.} In the following, we discuss the steps of our methodology in more detail.
\subsection{Data and estimation of cosmological parameters}
We constrain the parameters of each of the models presented in the last section using observations from {SLS}. We take into account only early-type galaxies as gravitational lenses, considering four measured properties: spectroscopically determined stellar velocity dispersion $\sigma$, the Einstein radius $\theta_E$, the lens redshift $z_l$, and the source redshift $z_s$. We use the fiduciary sample presented in \cite{Amante:2019xao} with a total of 143 systems,   {this sample includes systems with velocity dispersions in the range \(116 \, \mathrm{km/s} \leq \sigma \leq 342 \, \mathrm{km/s}\), lens redshifts in \(0.0625 \leq z_{\mathrm{l}} \leq 0.958\), source redshifts in \(0.2172 \leq z_{\mathrm{s}} \leq 3.595\), and Einstein radii in the range \(0.36'' \leq \theta_E \leq 2.55''\)}. In addition, we use a sub-sample {of} 99 systems discussed in \cite{Lizardo:2020wxw} which avoids nonphysical results when {calculating} the angular diameter distances in a Taylor series approximation, as {required} in cosmography\footnote{The 99 systems give a physical angular diameter distance in $y$-redshift, but not all of them give a physical result in $z$-redshift.},  {this sample contain measurements in the following ranges:  \(164 \, \mathrm{km/s} \leq \sigma \leq 326 \, \mathrm{km/s}\),  \(0.0625 \leq z_{\mathrm{l}} \leq 0.716\),  \(0.2172 \leq z_{\mathrm{s}} \leq 1.47\) and \(0.52'' \leq \theta_E \leq 1.78''\)}.

In the strong lensing phenomenon, the deflection of the light beam is so significant that it can lead to the formation of arcs, multiple images, or even Einstein rings. These images contain crucial properties that are valuable for modeling a lens system, thereby providing essential insights into various astrophysical phenomena. Consequently, observations of strong gravitational lensing play a pivotal role in analyzing the evolution of the Universe. Our approach involves a method that describes the mass distribution of the lens by correlating the separation between multiple images of the source with the angular diameter distances to the lens and to the source. If the lens model fitted to the observed images is assumed to be a spherical mass distribution, described by a singular isothermal sphere (SIS) model, the Einstein radius is defined as \cite{2006eac..book.....S} 
\begin{equation}
\theta_{E} = 4 \pi \frac{\sigma^2 D_{ls}}{c^2 D_{s}}, \label{thetaESIS}
\end{equation}
where $c$ represents the speed of light, $D_{s}$ denotes the angular diameter distance to the source, $D_{ls}$ stands for the angular diameter distance between the lens and source, and $\sigma$ represents the measured velocity dispersion of the lensing galaxy. The mass distribution of elliptical galaxies is well described as isothermal, as indicated by previous studies \cite{1995ApJ...445..559K,Grillo:2007iv,Treu:2002ee,Koopmans:2002qh,Rusin:2003hn}.

 {The angular diameter distance for any spatially flat cosmological model is defined as:}
\begin{equation}
D(z)=\frac{c}{H_0(1+z)}\int_0^z\frac{dz^{\prime}}{E(z^{\prime})},
\label{eq:D}
\end{equation}
\noindent where $H_{0}$ is the Hubble constant and
$E(z)$ is the dimensionless Friedmann parameter, presented in the previous section for the models of interest in this work. {By defining the ratio} $D \equiv D_{ls}/D_{s}$, the observational lens equation in the isothermal regime can be expressed as
\begin{equation}
D^{obs} = \frac{c^2 \theta_{E}}{4 \pi \sigma^2}. \label{Dlens}
\end{equation} 
The theoretical counterpart is defined as follows
\begin{equation}
D^{th}\left(z_{l}, z_{s}; \Theta \right) = \frac{\int^{z_{s}}_{z_{l}}\frac{\mathrm{dz}^{\prime}}{E(z^{\prime},\Theta)}}{\int^{z_{s}}_{0}\frac{\mathrm{dz}^{\prime}}{E(z^{\prime},\Theta)}}, \label{Dth}
\end{equation}
where $\Theta$ represents the free parameters for each cosmological model, $z_{l}$ is the redshift of the lens and $z_{s}$ is the redshift of the source. Therefore, we can constrain cosmological parameters minimizing the chi-square function,
\begin{equation}
\chi_{\mbox{SL}}^2(\Theta) = \sum_{i=1}^{N_{SL}} \frac{ \left[ D^{th}\left(z_{l}, z_{s}; \Theta \right)  -D^{obs}(\theta_{E},\sigma^2)\right]^2 }{ (\delta D^{\rm{obs}})^2},
\label{eq:chisquareSL}
\end{equation}
where $\delta D^{\rm{obs}}$ is the standard error propagation of the observational lens equation (\ref{Dlens}), $N_{SL}$ accounts for the number of strong lensing systems under consideration and $D^{th}$ is given by Eq.~\eqref{Dth}.

The posterior probability density function (PDF) of the free parameters $\mathbf{\Theta}$ for each model is estimated through a Markov chain Monte Carlo (MCMC) Bayesian statistical analysis. We {use} the Affine Invariant MCMC Ensemble sampler from the \texttt{emcee} Python module~\cite{2013PASP..125..306F} with 1000 burn-in steps, 1000 walkers, and 5000 MCMC steps. We estimate the cosmological parameters considering the likelihood function 
\begin{equation}
\mathcal{L}(\mathbf{\Theta})\propto e^{-\chi_{\mathrm{SL}}^{2}(\mathbf{\Theta})},
\end{equation}
where $\chi_{\mathrm{SL}}^{2}(\mathbf{\Theta})$ is given by Eq. (\eqref{eq:chisquareSL}). We assess chain convergence using the Gelman-Rubin test introduced by Gelman et al. (1992).  {The priors {for} the cosmological parameters  {of} each model are presented in Table \ref{tab:priors_models}.}
\begin{table}
    \centering
    \begin{tabular}{@{}|l|c|c|@{}}
    \hline
        {Models} & {Parameters} & {Priors} \\ \hline
        $\omega$CDM     & $\Omega_{{0m}}, \omega$          & $\Omega_{{0m}} \sim \mathcal{N}(0.3111, 0.0056),\; \omega \in [-4, 1]$ \\ \hline
        hFRW            & $\Omega_{0B}, \Omega_{0\lambda}$                         & $\Omega_{0B} \in [0, 1],\; \Omega_{0\lambda} \in [0, 1]$ \\ \hline
        DHOST           & $c_2, \beta$       & $c_2 \in [0.8, 1.2],\; \beta \in [-6, -2]$ \\ \hline
        $f(R,T)$          & $n$                   & $n \in [0, 3]$ \\ \hline
    \end{tabular}
    \caption{ {Priors used in cosmological models. $\mathcal{N}(\mu, \sigma)$ denotes a Gaussian distribution with mean $\mu$ and standard deviation $\sigma$. We assume a Gaussian prior on $\Omega_{0m}$ according to Planck measurements \cite{Planck:2018vyg}.}}
    \label{tab:priors_models}
\end{table}
To estimate the best fit values for the cosmological parameters we perform six runs employing two distinct data sets, one for each model (excluding the DHOST model). The DHOST model is addressed in a different way; since there is no analytical expression for $H(z)$ and for $D^{th}$, {various families of curves are constructed by varying the parameters $c_2$ and $\beta$. In the first case, we vary $c_2$ in the range $[0.8,1.2]$ with increments $\Delta c_2=0.02$, while keeping the remaining parameters fixed as $c_3 = 5, c_4=1, \beta=-5.3$. In the second case, $\beta$ is taken in the range $[-6,-2]$ with steps of $0.5$, and $c_2=1, c_3 = 5, c_4=1$. These choices are motivated by the results of~\cite{Crisostomi:2017pjs}. These curves are subsequently fitted using Eq.~(\ref{eq:chisquareSL}), and ultimately, the set of values corresponding to the solution that minimizes the likelihood function is chosen as the best fit for this model.}

\subsection{Direct estimation of cosmographic parameters}
The theoretical counterpart displayed in Eq.~\eqref{Dth} is highly dependent on the form of $E(z)$. However, we can avoid considerations of a specific model of gravity by using a parameterization of $E(z)$, such as the one obtained in the cosmographic scenario. A detailed analysis of this scenario in the context of {SLS} is presented in~\cite{Lizardo:2020wxw}, and the {bases are} explained in~\ref{appendix:a}. By using this, we obtain expressions for the distances between objects just in terms of their redshift and of derivatives of the scale factor, or the so called cosmographic parameters. Thus, in cosmography, Eq.~\eqref{Dth} is approximated
as

\begin{align}
    D^{th}\left(z_{l}, z_{s}; \Theta \right) & = \frac{\left(1 + z_s  \right)^{-1} \left[  \Delta(z_s) - \Delta(z_l) \right]}{D(z_s)} \nonumber \\
  & \approx \left(1-\frac{z_l}{z_s}\right)-\frac{(1+q_0)(z_s-z_l)z_l}{2 z_s} + \mathcal{O}[z^2],
    \label{Dthz}
\end{align}
where $\Delta(z_l), \Delta(z_s)$ are the  transverse comoving distances to the lens and source, respectively (see \cite{Hogg:1999ad, Lizardo:2020wxw} for more details).

 {In cosmography, the radius of convergence for any expansion in redshift  $z$ is limited to $ z = 1$, meaning that for  $z > 1$, such expansions may not reliably approximate cosmological models, including $\Lambda$CDM. To address this limitation, Cattoën and Visser ~\cite{Catto_n_2007} proposed the  $y$-redshift parameterization, defined as  $y = \frac{z}{1+z}$, which improves convergence properties and provides better constraints on cosmographic parameters when using Strong Lensing Systems (SLS) data. This approach has been demonstrated in studies such as Lizardo et al. \cite{Lizardo:2020wxw}, where the theoretical lens equation is expressed in terms of $y$  to enhance the analysis,}
\begin{align}
    D^{th}\left(y_{l}, y_{s}; \Theta \right)&= \frac{\left(1 - y_s  \right) \left[  \Delta(y_s) - \Delta(y_l) \right]}{D(y_s)} \nonumber \\
    & \approx \frac{y_s-y_l}{y_s-y_sy_l}-\frac{(1+q_0)(y_s-y_l)y_l}{2(y_s(-1+y_l)^2)} + \mathcal{O}[y^2],\label{Dthy}
\end{align}
where $\Delta(y_l), \Delta(y_s)$ are the transverse comoving distances to the lens and source in $y$-redshift space, $y_s=y(z_z)$ and $y_l=y(z_l)$. {With these results}, we can also constrain the cosmographic parameters using the previously defined chi-square function~\eqref{eq:chisquareSL}, where $D^{th}$ is now given by Eq.~\eqref{Dthz} when we want to constrain the
cosmographic parameters in $z$-redshift. For $y$-redshift the equation for chi-square is essentially the same, but replacing $D^{th}(z_l,z_s;\Theta)$ with
$D^{th}(y_l,y_s;\Theta)$
given by Eq.~\eqref{Dthy} and performing the appropriate transformations of the data and errors. We carried out two tests in the $y$-redshift space to estimate cosmographic parameters (one for each data set) assuming the following flat priors: $-0.95<q_0<-0.2$, $0<j_0<2.0$, $-2.0<s_0<7.0$ and $-5.0<l_0<10.0$.

\subsection{Assessing the applicability of cosmography} \label{reversalcosmo}
From Eq.~\eqref{eq:chisquareSL} we can obtain the best fit values for the cosmographic parameters according to {SLS}.  These values can be regarded as purely observational, and we refer to them as \textit{direct cosmographic parameters}. On the other hand, different sets of \textit{indirect} values for the cosmographic parameters can be obtained through the best fits for the free parameters of each model. These indirect parameters are obtained as follows. From the definition of the cosmographic parameters, we have
\begin{eqnarray}
    \frac{dH}{dz}\Big|_{z=0} & = & H_0 (1+q_0), \label{cosmo1}\\
    \frac{1}{2!}\frac{d^2H}{dz^2}\Big|_{z=0} & = &\frac{H_0}{2}(-q_0^2+j_0),\label{cosmo2}\\
    \frac{1}{3!}\frac{d^3H}{dz^3}\Big|_{z=0} & = &\frac{H_0}{6}(3q_0^2 + 3q_0^3 -4q_0j_0 - 3j_0 -s_0),\label{cosmo3}\\
    \frac{1}{4!}\frac{d^4H}{dz^4}\Big|_{z=0} & = &\frac{H_0}{24}(-12q_0^2 - 24q_0^3 - 15q_0^4 + 32q_0j_0 \nonumber \\ & & + 25q_0^2j_0 + 7q_0s_0 + 12j_0 - 4j_0^2 + 8s_0 + l_0).\label{cosmo4}
\end{eqnarray}
These equations can be inverted to give $q_0, j_0, s_0, l_0$ in terms of the derivatives of $H$ evaluated at $z=0$, 
\begin{eqnarray}
    q_0 & = & \frac{H'}{H_0}-1, \label{q0inv}\\
    j_0 & = & \frac{H''}{H_0}+\left(\frac{H'}{H_0}-1\right)^2, \label{j0inv}\\
    s_0 & = & \frac{H'''}{H_0}-3\left(\frac{H'}{H_0}-1\right)^2-3\left(\frac{H'}{H_0}-1\right)^3+\nonumber \\
    & & 4\left(\frac{H'}{H_0}-1\right)\left(\frac{H''}{H_0}+\left(\frac{H'}{H_0}-1\right)^2\right)+3\left(\frac{H''}{H_0}+\left(\frac{H'}{H_0}-1\right)^2\right) ,\label{s0inv}
\end{eqnarray}
where, just in this case, the prime  represents derivatives with respect to the redshift evaluated at $z=0$. 
{A similar but more cumbersome expression can be obtained for $l_0$.} 
For each model, these derivatives can be {evaluated} from Eqs.~\eqref{eq:ezw},~\eqref{2hfrw},~\eqref{eq:ezfrt} and from the numerical solution for $H$ in the case of DHOST, using the cosmological parameters adjusted through  {SLS}. If cosmography provides
a good representation of a model, we expect the 
direct and indirect values to be in agreement. As we show in the next section,
this agreement is reached within $2\sigma$ for all the models considered in this work (using the restricted sample).

 {In this study, we adopt a third-order approximation of cosmography as it provides a balance between accuracy and complexity in constraining cosmological parameters. Previous studies, \cite{Zhang:2016urt,Hu:2022udt}   have systematically compared different expansion methods over varying redshift ranges and concluded that third-order expansions effectively minimize the risk of divergence while maintaining sufficient precision for low-to-moderate redshift observations.}  

{Finally, we study the possibility of reconstructing the parameters of the models from the cosmographic parameters. To do so, we rewrite Eqs.~\eqref{cosmo1}-\eqref{cosmo4} expressing the left hand sides with the corresponding expressions for each model, thus forming a systems of equations that can be solved for the parameters of the model in terms of the cosmographic parameters. Plugging in the direct values of the cosmographic parameters we obtain indirect estimations for the parameters of each model, which are then compared to their direct values obtained from Eq.~\eqref{eq:chisquareSL}. Once again, we find agreement within $2\sigma$ when using the restricted SLS sample. This methodology can be useful for constraining parameters of modified gravity without performing a direct comparison with the observational data.}

\section{Results}
\label{sec:res}
\subsection{Cosmological constraints}
\begin{figure}
\includegraphics[scale=0.9]{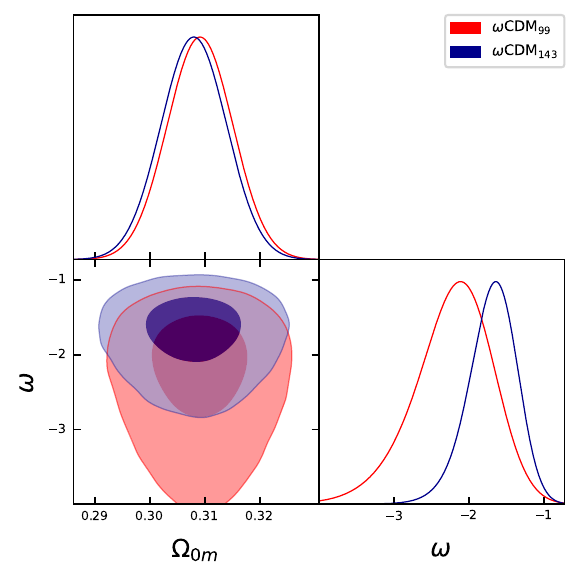}
\caption{1D marginalized posterior distributions and the 2D 68$\%$, 99.7$\%$ confidence levels for the $\Omega_{0m}$ and $\omega$ parameters for the $\omega$CDM model using a restricted sample of 99 SLS (red) and a sample of 143 SLS (dark blue). \label{fig:wCDM}}
\end{figure}

\begin{figure}
\includegraphics[scale=0.9]{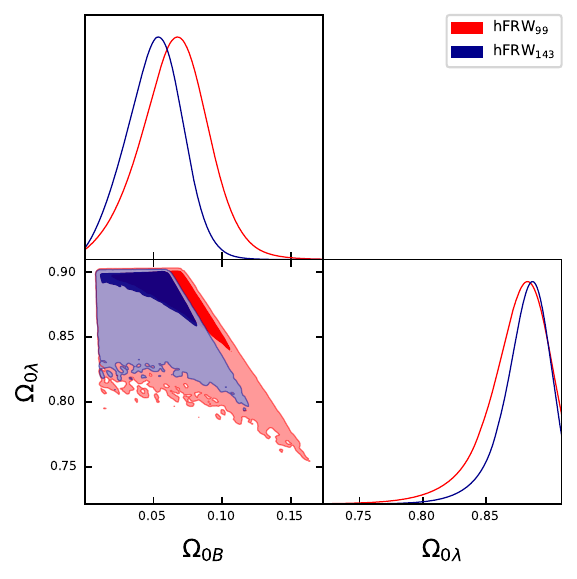}
\caption{1D marginalized posterior distributions and the 2D 68$\%$, 99.7$\%$ confidence levels for the $\Omega_{b}$ and $\Omega_{\lambda}$ parameters for the hFRW model using a restricted sample of 99 SLS (red) and a sample of 143 SLS (dark blue). \label{fig:Hfrw}}
\end{figure}

\begin{figure}
\includegraphics[width=7.5cm,scale=0.8]{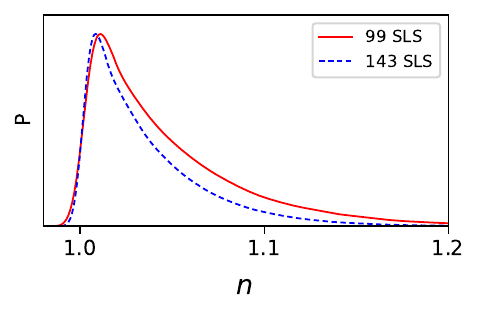}
\caption{1D marginalized posterior distributions for the $n$ parameter on the $f(R,T)$ model for a restricted sample of 99 SLS (red) and a sample of 143 SLS (blue).
\label{fig:FRT1}}
\end{figure}

The best fit values for the density parameters {of} each model are shown in the bottom panels of Tables~\ref{tab:99slsParam} and \ref{tab:143slsParam} for the restricted and fiduciary samples respectively. Additionally, we present the values of the {indirect} cosmographic parameters corresponding to each model through equations \eqref{cosmo1}, \eqref{cosmo2}, \eqref{cosmo3}, and \eqref{cosmo4}. For the cosmological constraints we {find} similar values with both samples, which are consistent at $1\sigma$ in all cases. In the $\omega$CDM model (see Figure \ref{fig:wCDM} to visualize the confidence contours and PDF) the value for $\omega$ that we find is within $1\sigma$ of the value reported by Planck ($\omega = -1.56^{+0.60}_{-0.48}$)~\cite{Planck:2018vyg}; however, it is only consistent at $2\sigma$ with a cosmological constant. It is noteworthy that the value of $\omega$ suggests the existence of a phantom dark energy in both cases. For the model hFRW (see Figure \ref{fig:Hfrw} to visualize confidence contours and PDF) {both samples lead to values of $\Omega_{0B}$ that are consistent at $1\sigma$ with those obtained in \cite{Cai:2018ebs} using supernova measurements}, indicating lower values for $\Omega_{0B}$ $\sim 6\%$ than those of the $\Lambda$CDM model ($\Omega_{0m} \sim 30 \%$) { where the inclusion of dark matter is required}. {On the other hand}, $\Omega_{0\lambda}$ is consistent with~\cite{Cai:2018ebs} at 2$\sigma$. In the $f(R,T)$ scenario (see figure \ref{fig:FRT1}) we obtain $n=1.033\pm0.05$ and $n=1.025\pm0.04$ for the restricted and fiduciary samples respectively, being consistent with a universe undergoing accelerated expansion as indicated in \cite{Nagpal:2018mpv}. Finally, for the DHOST model arising from a numerical approximation, we obtain $c_2=1.10$ and $c_2=1.08$ for the restricted an fiduciary samples, respectively, slightly greater than a canonical kinetic term ($c_2=1$)\footnote{Although $c_2$ can be absorbed in a redefinition of the scalar field, this would modify the other coupling constants. }. The error is taken as 0.02, which was the variation of the parameter $c_2$ among the different curves shown in Figure \ref{fig:hzdhost}.

\begin{table}
    \centering
    \resizebox{\columnwidth}{!}{
    \begin{tabular}{|c|c|c|c|c|c|}
        \hline
        Param. & hFRW & $\omega$CDM & $f(R,T)$ & DHOST & Cosmo Obs. \\ \hline
        $q_0$ & $-0.84\pm0.06$ & $-1.7\pm0.8$ & $-0.484\pm 0.025$ & $-0.722\pm0.009$ & $-0.54\pm0.1$\\ \hline
        $j_0$ & $1.0\pm1.3$ & $8\pm7$ & $0.25\pm0.08$ & $0.475\pm0.027$ & $0.36\pm0.52$ \\ \hline
        $s_0$ & $0.0\pm692$ & $37\pm79$  & $0.1\pm0.4$ & $1.42\pm0.24$ & $3.7\pm3.31$ \\ \hline
        $l_0$ & $0.0\pm5000$ & $-600\pm1600$ & $-0.6\pm3.1$ &  & $-0.35\pm2.22$ \\ \hline
        \hline
        \multirow{2}{*}{Free Param.} & $\Omega_{0B}=0.068\pm0.025$ & $\Omega_{0m}=0.31\pm0.005$  & \multirow{2}{*}{$n=1.033\pm0.05$} & \multirow{2}{*}{$c_2 = 1.10\pm0.02$} & \\
        & $\Omega_{0\lambda}=0.883\pm0.027$  & $\omega=-2.134\pm0.52$ &   &  & \\ \hline
    \end{tabular}
    }
    \caption{Values for the cosmographic parameters obtained for each of the models of MG analyzed in the present work. The values in the bottom panel of each model are the density parameters obtained from the SLS data for each of the models through Bayesian statical analysis. For both cases, here we used the sample of 99 SLS. The DHOST data are only computed up to the third parameter, as the numerical approximation lacks sufficient precision to obtain a solution up to the fourth order.}
    \label{tab:99slsParam}
\end{table}

\begin{table}
    \centering
    \resizebox{\columnwidth}{!}{
    \begin{tabular}{|c|c|c|c|c|c|}
        \hline
        Param. & hFRW & $\omega$CDM & $f(R,T)$ & DHOST & Cosmo Obs. \\ \hline
        $q_0$ & $-0.83\pm0.05$ & $-1.2\pm0.5$ & $-0.488\pm 0.02$ & $-0.713\pm0.01$& $-0.51\pm0.06$\\ \hline
        $j_0$ & $1.0\pm0.9$ & $4.3\pm3.5$ & $0.25\pm0.06$ & $0.462\pm0.031 $& $0.16\pm0.24$ \\ \hline
        $s_0$ & $0.0\pm418$ & $12\pm32$  & $0.11\pm0.29$ & $1.5\pm0.16$ & $4.69\pm2.5$ \\ \hline
        $l_0$ & $0.0\pm3200$ & $-147\pm464$ & $-0.6\pm3.1$ &  & $-0.37\pm0.45$ \\ \hline
        \hline
        \multirow{2}{*}{Free Param.} & $\Omega_{0B}=0.054\pm0.02$ & $\Omega_{0m}=0.308\pm0.006$  & \multirow{2}{*}{$n=1.025\pm0.04$} &  \multirow{2}{*}{$c_2=1.08\pm0.02$}  & \\
        & $\Omega_{0\lambda}=0.886\pm0.02$  & $\omega=-1.653\pm0.322$ &  & & \\ \hline
    \end{tabular}
    }
    \caption{Values for the cosmographic parameters obtained for each of the models of MG analyzed in the present work. The values in the bottom panel of each model are the density parameters predicted from the SLS data for each of the models. For both cases, here we used the sample of 143 SLS. The DHOST data are only computed up to the third parameter, as the numerical approximation lacks sufficient precision to obtain a solution up to the fourth order.}
    \label{tab:143slsParam}
\end{table}

\subsection{Constrast with cosmography}
In the top panels of Tables \ref{tab:99slsParam} (99 SLS) and \ref{tab:143slsParam} (143 SLS) we show the indirect cosmographic parameters associated to each model. In addition, in the last column of each table we show the direct cosmographic parameters obtained in the $y$-redshift space for both data sets~(similar values are reported in~\cite{Lizardo:2020wxw}).
If we compare the values of the cosmographic parameters of each model at the present time with those arising directly from SLS measurements using cosmography, we can see that the $q_0$ value is consistent at 1$\sigma$ only for the value given by the $f(R,T)$ model, the remaining models are only consistent at 2$\sigma$ of the confidence level, excluding the DHOST and hFRW models for the fiduciary sample, which only exhibits consistency at 3$\sigma$.  It is worth noticing that all models are compatible with accelerated expansion. Regarding the parameter $j_0$, most of the models are in agreement at 1$\sigma$ of the confidence level with the one obtained directly from cosmography, except for the $\omega$CDM model and the DHOST model analysed with the fiduciary sample, which are consistent at 2$\sigma$. For the remaining two parameters $s_0$ and $l_0$, the majority are consistent at 1$\sigma$ of confidence level with those obtained directly from cosmography, except for models $f(R,T)$ and DHOST (using the fiduciary sample), which demonstrate consistency up to 2$\sigma$.  {Regarding the compatibility with $\Lambda$CDM,  as reported in \cite{Lizardo:2020wxw}  the cosmographic parameters associated to the $\Lambda$CDM scenario are $q_0=-0.525\pm0.013$, $j_0=1$, $s_0=-0.42\pm0.04$ and $l_0=2.401\pm0.035$. With these under consideration we can see that the cosmographic parameters computed for each of the models are, in general, 
far from the ones expected in the $\Lambda$CDM scenario, although for all models, including $\Lambda$CDM, the results are within 3$\sigma$ of the direct cosmographic parameters. This highlights the usefulness of cosmography as an umbrella parametrization, i.e., a parametrization that covers a wide range of models.} Furthermore, it is observed that the errors in the cosmographic parameters tend to increase as more orders of expansion are considered. This growth pattern aligns with the observations reported in studies using SN Ia data, as exemplified by \cite{Zhang_2017}.

\begin{figure}
    \centering
\includegraphics[width=0.49\textwidth]{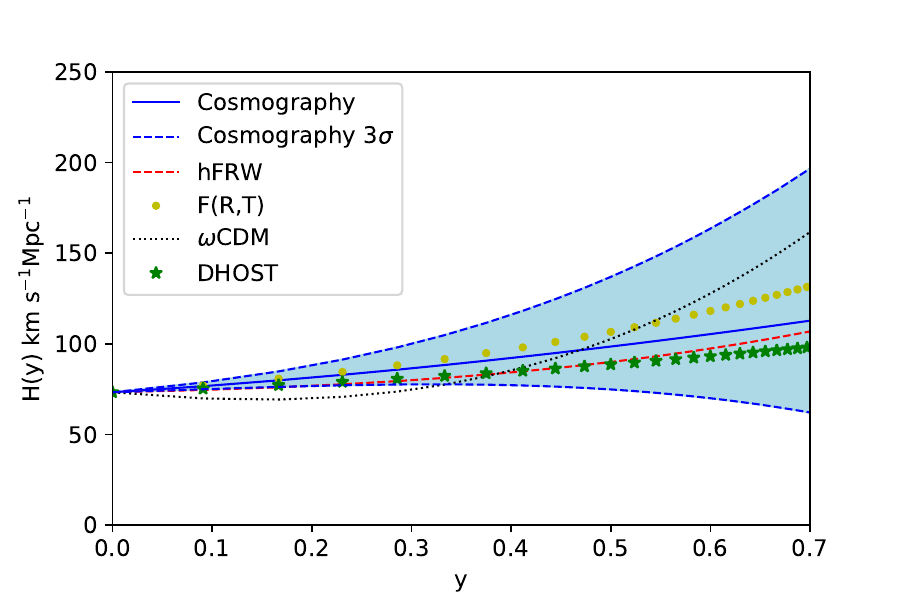} \includegraphics[width=0.49\textwidth]{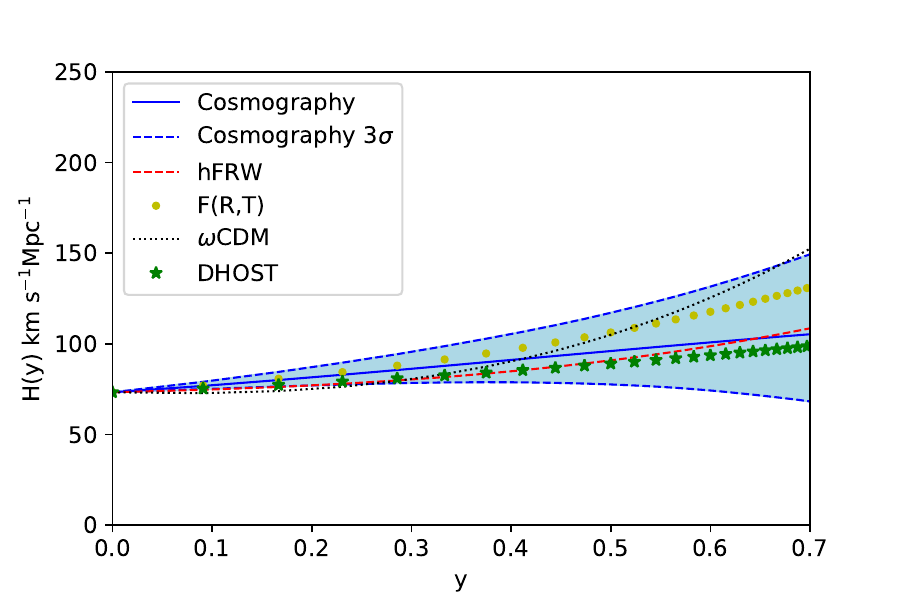}
    \caption{Comparison between $H(y)$ obtained with cosmographic parameters inferred from each model and the one obtained directly from cosmography with 99 (left) and 143 (right) SLS. The minimum displayed by $H(z)$ in the $\omega$CDM is related to the violation of the null energy condition (see, e.g.~\cite{Yang:2019vgk}).}
    \label{HyTODAS99}
\end{figure}

{In order to get some insight into the meaning of the values presented above, we take the comparison to the level of the Hubble factor. For each model, we construct an \textit{indirect cosmographic Hubble factor} by evaluating Eq.~\eqref{Hubbley} with the corresponding indirect cosmographic parameters. In Fig.~\ref{HyTODAS99}, we compare these indirect Hubble factors to the one obtained with the direct cosmographic parameters. We also show the $3\sigma$ confidence region of the direct Hubble factor, obtained by propagating the Markov chains of the statistical analysis.} As a quantitative measure of the difference between direct and indirect Hubble factors, we use the mean relative error. 
We found similar results for both samples, with the hFRW model showing the minimum deviation from the true model, obtaining 6.7$\%$ and 3.6$\%$ for restricted and fiduciary samples, respectively. The second model with the smallest difference was the DHOST model, which produced a disparity of 9.5$\%$ for the restricted sample and 6$\%$ for the fiduciary sample. Subsequently, the model $f(R,T)$ secured the third position, showing a discrepancy of 9.6$\%$ and 13.2$\%$, respectively. Finally, the model with a larger scatter was $\omega$CDM, which yielded a relative error of approximately $20\%$ for both data sets. Since each cosmological model has a different number of free parameters, in the following section, we present an analysis that takes this into consideration.

\subsection{Model selection}
{As a final test of the compatibility between cosmography and the models of modified gravity under consideration, we use model selection criteria. Specifically, we investigate whether these criteria give consistent results when applied either to the models fitted directly with SLS or to their indirect cosmographic approximations.}
Model selection aims to strike a balance between the accuracy of fitting observational data and the predictiveness of the model that achieves such a fit. This balance is attained through {a criterion that} assigns a numerical value to each model, enabling the creation of a rank-ordered list. We employ the Bayesian information criterion (BIC) \cite{Schwarz:1978tpv} and the Akaike information criterion (AIC) \cite{Akaike:1974}. These criteria are defined by
\begin{eqnarray} 
\mbox{BIC}&=&\chi^{2}_{min} + k \ln N, \label{BIC}\\ 
\mbox{AIC}&=&\chi^{2}_{min} +2k\label{AIC},
\end{eqnarray}
where $N$ stands for the number of data points, $k$ is the number of parameters and $\chi^{2}_{min}$ is the chi-square function.  When these criteria are applied to the models fitted directly with SLS, $\chi^{2}_{min}$ is evaluated with the best fit cosmological values from the bottom panels of tables \ref{tab:99slsParam} and \ref{tab:143slsParam}. On the other hand, when the criteria are applied to the cosmographic approximations, $\chi^{2}_{min}$ is evaluated using the   theoretical lens equation \eqref{Dthy} expressed in terms of the cosmographic parameters in $y$-redshift space, and taking the indirect cosmographic parameters from the corresponding columns of tables \ref{tab:99slsParam} and \ref{tab:143slsParam}. Tables \ref{modelselection99} and \ref{modelselection143} show the values of the selection criteria for both approaches. Furthermore, we present the reduced chi-square $\chi^2_{red} = \chi^2_{min} / (N - k) $ and the $\Delta\mathrm{BIC} = \mathrm{BIC}_{i} - \mathrm{BIC}_{min}$ ($\Delta\mathrm{AIC} = \mathrm{AIC}_{i} - \mathrm{AIC}_{min}$), which is a relative value of BIC (AIC) for model i compared to the minimum $AIC_{min}$ ($BIC_{min}$) {among} all models (see \cite{Shi:2012ma} for details). Considering the current datasets, the hFRW model is favored by all model selection tests; however, for the sample of 99 SLS, the preferred model (in the cosmographic approach) is the DHOST model; nevertheless, the values for $\Delta\mathrm{BIC}$, $\Delta\mathrm{AIC}$, and $\chi^2_{red}$ are very close to those obtained for the hFRW model, showing not enough evidence to discern between both models. Furthermore, the least accepted model was the $\omega$CDM model. 
This could be due to the Gaussian prior (from Planck measurements) on $\Omega_{0m}$,  which appears to impact the model's compatibility with SLS. Assuming such a prior is necessary due to the
inability to properly constrain the value of $\Omega_{0m}$ with SLS {at galactic scales}~\cite{Chen:2018jcf,Amante:2019xao}.  {We also attribute these deviations to the reconstructed $H(z)$ profile for $\omega$CDM model, which shows a minimum not seen in the other models (see Figure \ref{HyTODAS99}). This minimum occurs close to the expansion point ($y = 0$), potentially amplifying differences in cosmographic parameters, leading to elevated AIC and BIC values. Future research could extend this analysis with higher-order expansions or alternative data sets to explore these discrepancies further.} It is worth noticing that departures from $\Lambda$CDM {using measurements at galaxy cluster scales} have been previously found in \cite{Caminha:2021iwo,Limousin:2022lvv}, and more recently, employing observations from both scales in~\cite{Verdugo:2024dgs}.  {Deviations from
$\Lambda$CDM predictions have also been reported recently by the DESI collaboration \cite{DESI:2024hhd} when using the CPL parameterization of dark energy. } 
\begin{table}
    \centering
    \resizebox{\columnwidth}{!}{
    \begin{tabular}{|c|c|c|c|c|c|c|c|}
         \hline
         Model & Free parameters & $\chi^2_{SLS}$ & $\chi^2_{red}$ & AIC &  BIC & $\Delta$AIC & $\Delta$BIC \\  \hline
          $hFRW$ & 2 & 177.624 &  1.831 &181.624 & 187.550 &  0 & 0 \\ \hline
         $\omega$CDM & 2 & 192.773 & 1.987 & 196.773 & 201.963  &  15.149 & 14.413 \\ \hline
         $f(R,T)$ & 1 & 190.978 & 1.948 & 192.978 & 195.573 & 11.354  & 8.023 \\ \hline
         DHOST & 1 & 190.326 & 1.942 & 192.326 & 194.921 & 10.702  & 7.371 \\ \hline
         \hline
          $hFRW_C$ & 2 & 121.255 & 1.250 & 125.255 & 130.445 & 1.104  & 3.699\\ \hline
        $\omega$CDM$_C$ & 2 & 620.986 &  6.40 & 624.986 & 630.176 & 500.835  & 503.436 \\ \hline
         $f(R,T)_C$ & 1 & 132.845 & 1.355 & 134.845 & 137.440 & 10.334  & 10.694  \\ \hline
         DHOST$_C$ & 1 & 122.151 & 1.246 & 124.151 & 126.746 & 0  & 0 \\ \hline
    \end{tabular}
    }
    \caption{Model selection criteria for the restricted sample with a total of 99 SLS. The subscript C indicates that the chi-square function \eqref{eq:chisquareSL} was constructed using the cosmography parameters to a third-order approximation of each cosmological model and extracted from table \ref{tab:99slsParam}.}
    \label{modelselection99}
\end{table}

\begin{table}
    \centering
    \resizebox{\columnwidth}{!}{
    \begin{tabular}{|c|c|c|c|c|c|c|c|}
         \hline
         Model & Free parameters & $\chi^2_{SLS}$ & $\chi^2_{red}$ & AIC &  BIC & $\Delta$AIC & $\Delta$BIC \\  \hline
         $hFRW$ & 2 & 231.103 & 1.639 & 235.103 & 241.029 &  0 & 0\\ \hline
         $\omega$CDM & 2 & 263.795 & 1.871 & 267.795 & 273.721  &  32.692 & 32.692\\ \hline
         $f(R,T)$ & 1 & 243.121 & 1.712 & 245.121 & 247.716 & 10.018  & 6.687 \\ \hline
         DHOST & 1 & 253.170 & 1.783 &255.170 & 258.133 & 20.067  & 17.104 \\ \hline
          \hline
          $hFRW_C$ & 2 & 168.139 & 1.192 & 172.139 & 178.065 &  0 & 0\\ \hline
        $\omega$CDM$_C$ & 2 & 436.528 &  3.096 & 440.528 & 446.453  & 268.389  & 268.388\\ \hline
         $f(R,T)_C$ & 1 & 193.435 & 1.362 & 195.435 & 198.398  & 23.296 & 20.333  \\ \hline
         DHOST$_C$ & 1 & 179.782 & 1.266 & 183.782 & 184.745 & 11.643  & 6.68 \\ \hline
    \end{tabular}
    }
    \caption{Model selection criteria for the fiduciary sample with a total of 143 SLS. The subscript C indicates that the chi-square function \ref{eq:chisquareSL} was constructed through the cosmography parameters to a third-order approximation of each cosmological model extracted from table \ref{tab:143slsParam}.}
    \label{modelselection143}
\end{table}
\subsection{Reconstruction}

\begin{figure}
    \centering
    \includegraphics[width=0.9\columnwidth]{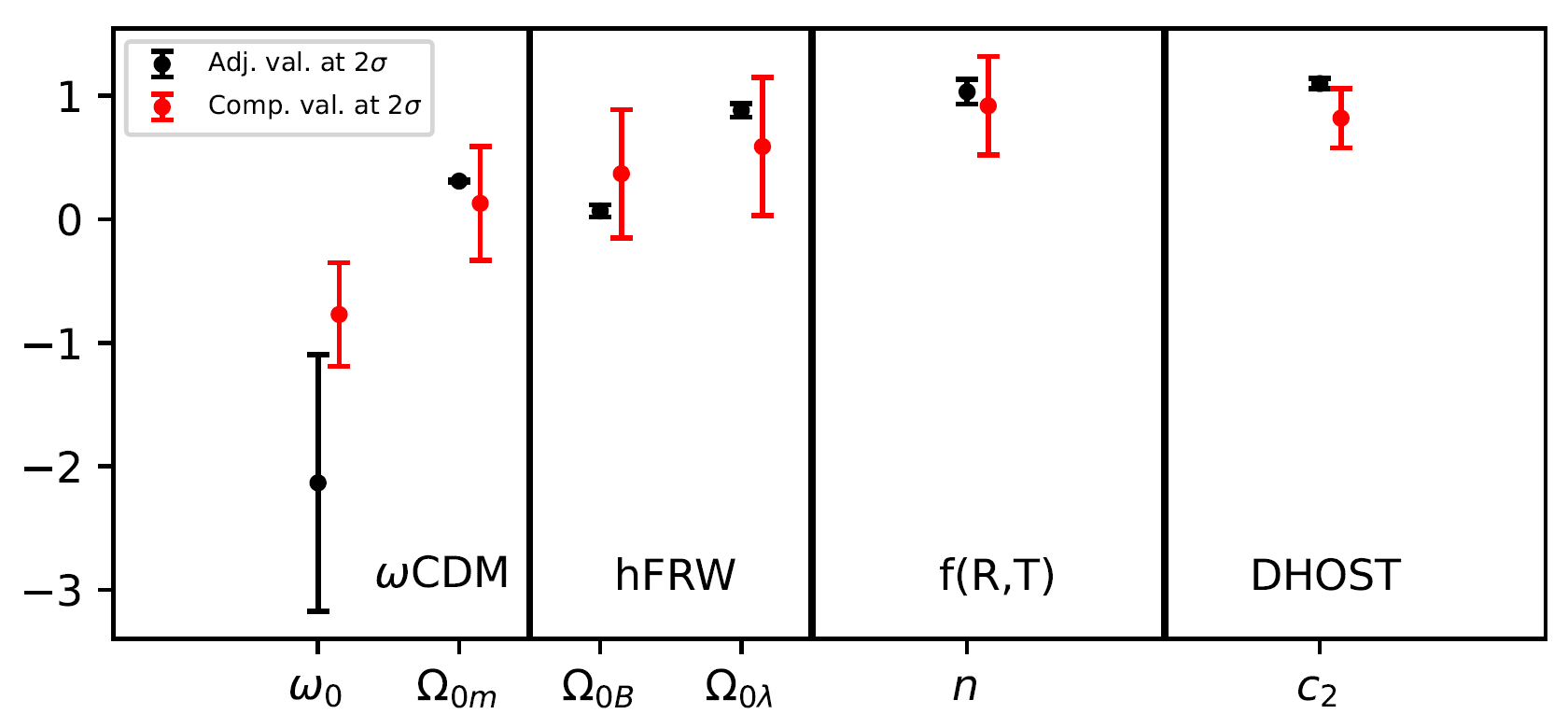}
\caption{Comparison of the density parameters of each cosmological model. In black we show the values obtained by direct fitting of the model to observational data, while in red we show the values computed from the cosmographic parameters. Here, we use the 99 SLS sample 
(see upper-right column of table  \ref{tab:99slsParam}). The values shown in red were computed using the process described in the final paragraph of section \ref{reversalcosmo}. 
}
    \label{fig:densitycomparison99}
\end{figure}

\begin{figure}
    \centering
    \includegraphics[width=0.9\columnwidth]{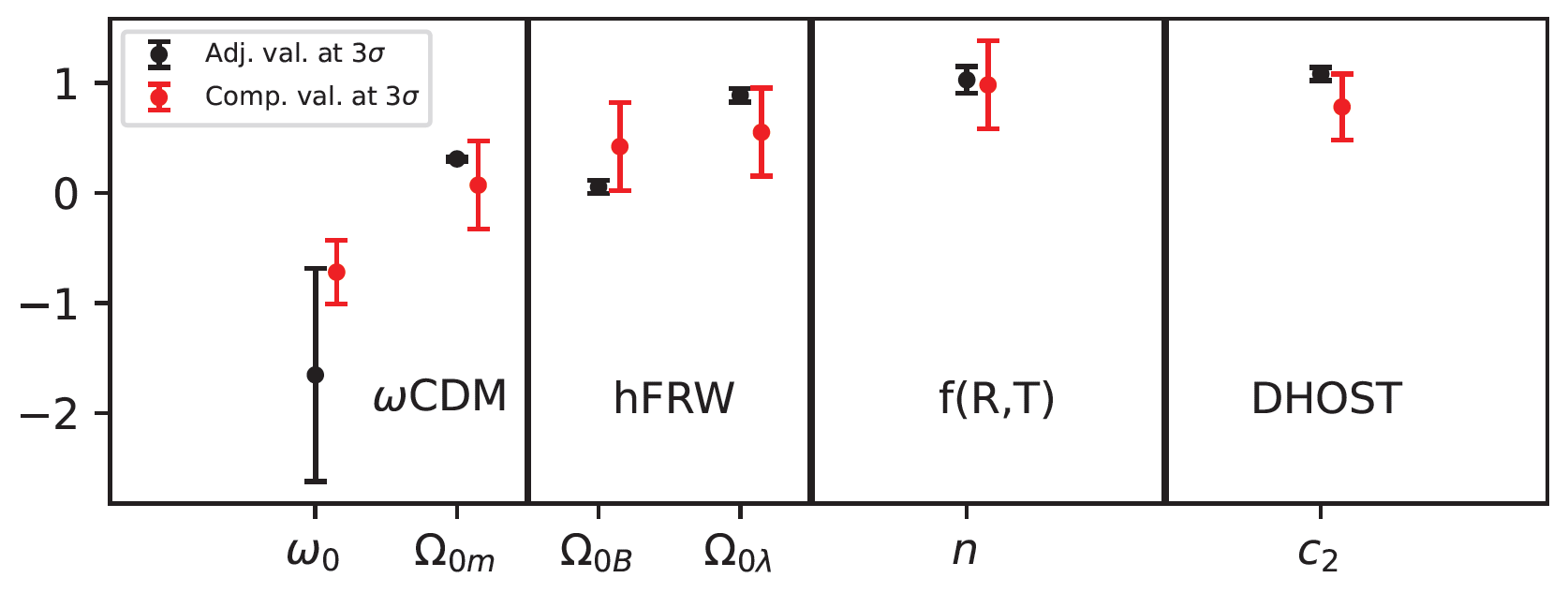}
\caption{Comparison of the density parameters of each cosmological model. In black we show the values obtained by direct fitting of the model to observational data, while in red we show the values computed from the cosmographic parameters. Here, we use the 143 SLS sample 
(see upper-right column of table  \ref{tab:143slsParam}). The values shown in red were computed using the process described in the final paragraph of section \ref{reversalcosmo}. 
}
    \label{fig:densitycomparison143}
\end{figure}

\begin{table}
    \centering
    \begin{tabular}{|c|c|c|}
         \hline
         Model & 99 SLS & 143 SLS  \\ \hline
         $f(R,T)$  & $n=0.92\pm0.2 $& $n = 0.98\pm0.12$\\ \hline
         \multirow{2}{*}{$\omega$CDM} & $\Omega_{0m} = 0.13\pm0.23$ & $\Omega_{0m} = 0.07\pm0.13$ \\
         & $\omega_0 = -0.77\pm0.21$ & $\omega_0 =-0.72\pm0.11 $\\ \hline
         \multirow{2}{*}{hFRW} & $\Omega_{0B} = 0.37\pm0.26$ & $\Omega_{0B} = 0.42\pm0.12$ \\
         & $\Omega_{0\lambda} = 0.59\pm0.28$ & $\Omega_{0\lambda} = 0.55\pm0.13$\\ 
         \hline
         DHOST & $c_2 = 0.82\pm0.12$ & $c_2 = 0.78\pm0.10$\\
         \hline
    \end{tabular}
    \caption{The values presented here are the values for the density parameters of each cosmological model computed directly from the cosmographic parameters obtained from direct observations of the SLS data. Errors are calculated at 1$\sigma$ of the confidence level.}
    \label{reversetab}
\end{table}
{So far we have verified in detail the compatibility between cosmography and some models of modified gravity emerging from diverse theoretical scenarios. This supports the idea that one should be able to predict the best fit for the cosmological parameters of MG models using cosmography, i.e. without the need of a direct fit between the models and observational data. This is what we explore in this section.  }
We reverse the procedure from equations (\ref{cosmo1}-\ref{cosmo4}), now using the cosmographic parameters obtained directly from SLS to compute the density parameters of each model (see table \ref{reversetab}). {These values are to be compared with the direct fits provided in Tables \ref{tab:99slsParam} and \ref{tab:143slsParam}.} In the $\omega$CDM scenario, we find a discrepancy {beyond} $2 \sigma$ in the estimation of  $\omega$ {when using the sample with 143 SLS}. {Similar results are obtained for the other models, although with smaller uncertainties. Figures \ref{fig:densitycomparison99} and \ref{fig:densitycomparison143} illustrate the comparison between direct and indirect values of the cosmological parameters. For the reduced sample of 99 SLS, all the parameters are compatible at $2\sigma$. Therefore, we recommend the use of the restricted sample for estimating cosmological parameters.  { Although the use of cosmographic parameters to infer density parameters for a given modified gravity model may be limited in certain cases, their utility lies in providing a preliminary, fast comparison that, as we see from our results, lies within 2$\sigma$ of the actual value of the parameter's of each model. Incorporating additional observations could help narrow the gap.
}}

{As the previous discussion shows, the reverse procedure gives a reasonable estimation of the parameters of each model. Nevertheless, we see two issues:}
\begin{description}
    \item[Error propagation] {The complexity of the algebraic expressions resulting from solving (\ref{cosmo1}-\ref{cosmo4}) for the parameters of the model leads to large uncertainties due to error propagation. This can be reduced by improving the error bars of the cosmographic parameters with the inclusion of more data sets.}
    \item[Over-determined system] {While there is an arbitrary number of cosmographic parameters (depending on the  order of approximation), there is only a finite and fixed number of parameters of each model.} Therefore, the process is over-determined, since only a subset of equations (\ref{cosmo1}-\ref{cosmo4})  needs to be used, allowing for different values to be obtained in the estimation of the density parameters. We suggest employing the necessary equations in ascending order since lower-order terms are more relevant in cosmography, and also to avoid even larger error propagation -- lowest order expressions are usually simpler. 
\end{description}
Summarising, it is possible to use the reverse process described here in order to get a quick estimation of the parameters of a MG model, but it is essential to take into account the uncertainties of the cosmographic parameters and the resulting error propagation, and to interpret the results as a range where the model is compatible with the set of observations used to derive the cosmographic parameters.

\section{Discussion}
\label{sec:dis}

The main objective of this work was to analyze a methodology that allows for a simple and fast comparison between predictions of modified gravity and direct results from observations. The simplicity of this methodology relies on the fact that one does not need to evaluate the observables in the modified theory of gravity; instead, one only needs a numerical or analytic expression for the Hubble parameter of the theory. With this expression at hand, one can find a relation between the parameters of the theory and a kinematical parameterization of the properties of the Universe, called cosmography. 

In order to assess the validity of this methodology, we use a two-step process. First, we obtain the best fit parameters of the model, taking into consideration strong lensing observations in which an elliptical galaxy acts as a lens. With these parameters, we estimate indirect values for the cosmographic parameters and compare them to the direct best fits for cosmography obtained with the same data set. This step gives us information on the compatibility between the models and cosmography. In most cases we find good agreement within 2$\sigma$ of confidence level. As a complement to this step, we compare the 3th order cosmographic expressions for $H(y)$ evaluated with the direct and indirect cosmographic parameters. The smallest mean relative difference was for hFRW, showing a value less than 7$\%$ for both data sets. As a further analysis in this first step, we subjected the models to different statistical selection tests. The results of these tests are consistent with each other, indicating that the preferred model was hFRW. This result should be taken with caution, as the true model should be favored by all astrophysical observations; therefore, it is necessary to subject this model to more sets of observational data.
 
 Now, the second step is to determine quantitatively whether cosmography can be used to reconstruct the best fit parameters of each model. From a mathematical point of view, this is not straightforward, since the system of equations may be over-determined. For example, if a model contains only one parameter, then Eqs.~(\ref{cosmo1}-\ref{cosmo4}) can be manipulated to obtain 4 different equations for the parameters given in terms of given values of $q_0, j_0, s_0$ and $l_0$. Since in cosmography these values are independent, the four equations lead, in principle, to four different values for the only parameter of the model. The reasonable choice then is to start from the lowest orders and solve until we obtain values for all the parameters of the model. In this way, we find that the parameters of the model can indeed be reconstructed within $2\sigma$ of confidence level.

It is important to note that this work was conducted assuming a Maclaurin series, i.e., around $z=0$ ($y=0$) which corresponds to the present day of the Universe. Therefore, the accuracy decreases away from that point, as can be observed in Fig.~\ref{HyTODAS99}. {Alternatively}, the expansion could be  around another point to obtain accurate information {about different} stages of the Universe, for example, at the transition redshift or even in the early Universe. 
{This underscores the necessity of exploring, in future research, different approaches to cosmography and diverse astrophysical observations at different epochs in the Universe, which can lead to more robust and stronger constraints.}

 In conclusion, this study underscores the importance of investigating the relations among various cosmological models and cosmography. Understanding these relationships yields practical applications to modified gravity scenarios -- even if the solutions are known only numerically -- {such as the possibility} to probe the validity of {MG} with observational data in an  {efficient} manner.
\subsection*{Acknowledgments}

AL is supported by CONAHCyT-855158. MHA acknowledges support from \textit{Estancias Posdoctorales CONAHCyT}. JC is partially supported by DCF-320821.

\subsection*{Data Availability}
The data underlying this article were accessed from https://doi.org/10.1093/mnras/sty260, also https://doi.org/10.1093/mnras/staa2760. The criteria for the SLS data samples is taken from https://doi.org/10.1093/mnras/stab2465, and references therein.

\section*{References}

\bibliographystyle{unsrt}
\bibliography{main}

\appendix
\section{Cosmography in the $y$-redshift scenario}\label{appendix:a}
The foundations of this work are placed in a technique known as cosmography.  {It provides a model-independent framework for analyzing the kinematics of the Universe, enabling the derivation of cosmological insights directly from observational data without reliance on specific underlying models. Its utility has been demonstrated in constraining spatial curvature through combined datasets \cite{Li}, testing the cosmological principle using the Pantheon+ supernova sample \cite{Hu:2024qnx}, and addressing the Hubble constant tension via dipole–monopole corrections \cite{Hu} By directly relating observable quantities to the Universe’s expansion history, cosmography has proven to be a robust tool in modern cosmological investigations.}

For a detailed formulation we start with the scale factor $a(t)$ and expand it as a Taylor series around $t=t_0$
\begin{equation}
    a(t) = a(t_0) + \dot{a}(t_0)(t-t_0)-\frac{1}{2}\ddot{a}(t_0)(t-t_0)^2 + \dots ,
\end{equation}
and from here we divide the expression by $a(t_0)$ in order to obtain
\begin{equation}
    \frac{a(t)}{a(t_0)} = 1 + H_0(t-t_0) - \frac{q_0}{2}H_0^2(t-t_0)^2 + \dots , \label{taylor1}
\end{equation}
where
\begin{equation}
    q_0 \equiv -\frac{\ddot{a}(t_0)a(t_0)}{\dot{a}(t_0)^2},
\end{equation}
is the so called \textit{deceleration parameter}. As we can see, this parameter is dimensionless, and it is directly proportional to the second derivative of the scale factor. The suffix $0$ stands for present time parameters. Similarly, we define parameters related to higher order derivatives, which are named \textit{cosmographic parameters},
\begin{eqnarray}
     q & \equiv & -\frac{1}{a}\frac{d^2a}{dt^2}\left[\frac{1}{a}\frac{da}{dt}\right]^2, \label{qq}\\
     j & \equiv & \frac{1}{a}\frac{d^3a}{dt^2}\left[\frac{1}{a}\frac{da}{dt}\right]^3, \label{jerk}\\
     s & \equiv & \frac{1}{a}\frac{d^4a}{dt^2}\left[\frac{1}{a}\frac{da}{dt}\right]^4, \label{snap}\\
    l & \equiv & \frac{1}{a}\frac{d^5a}{dt^2}\left[\frac{1}{a}\frac{da}{dt}\right]^5,
    \label{lerk}
\end{eqnarray}
where $j$ is known as \textit{jerk}, $s$ is called \textit{snap} and $l$ is named \textit{lerk parameter}. Then equation (\ref{taylor1}) can be expressed as,
\begin{eqnarray}
    \frac{a(t)}{a(t_0)} & = & 1 + H_0(t-t_0) - \frac{1}{2}q_0H_0^2(t-t_0)^2 + \frac{1}{6}j_0H_0^3(t-t_0)^3\\ & & + \frac{1}{24}s_0H_0^4(t-t_0)^4 + \frac{1}{120}l_0H_0^5(t-t_0)^5 + \mathcal{O}(t^6).
\end{eqnarray}
 With a similar procedure we expand the Hubble factor into a Taylor series in terms of the redshift, and by using the previous definitions of the cosmographic parameters (equations \ref{qq}, \ref{jerk}, \ref{snap} and \ref{lerk} ) we find an algebraic expression in the following way:
\begin{eqnarray}
    \nonumber H(z) & = & H_0 + \frac{dH}{dz}\Big|_{z=0}z + \frac{1}{2!}\frac{d^2H}{dz^2}\Big|_{z=0}z^2 
  +
    \frac{1}{3!}\frac{d^3H}{dz^3}\Big|_{z=0}z^3 + \dots \\ \nonumber
     & = & H_0 \Big[ 1 + (1+q_0)z +\frac{1}{2}(-q_0^2 + j_0)z^2
     + \frac{1}{6}(3q_0^2 + 3q_0^3 -4q_0j_0 - 3j_0 -s_0)z^3 
     \\ \nonumber 
     & \ & + \frac{1}{24}(-12q_0^2 - 24q_0^3 - 15q_0^4 + 32q_0j_0 + 25q_0^2j_0 + 7q_0s_0 + 12j_0 \nonumber \\
     & \ & - 4j_0^2 + 8s_0 + l_0)z^4  +\mathcal{O}(z^5) \Big]. \label{Hubbletaylor}
\end{eqnarray}

\begin{figure}[h]
    \centering
    \includegraphics[width=0.8\textwidth]{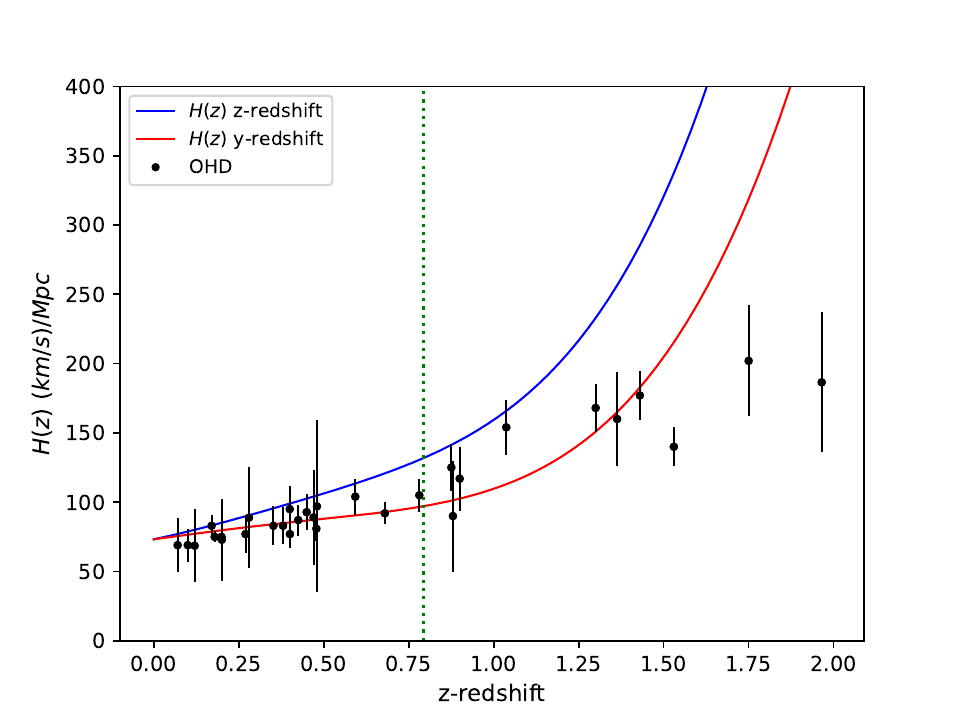}
    \caption{$H(z)$ in terms of the cosmographic parameters. The blue line represents the equation \ref{Hubbletaylor} in terms of cosmographic parameters obtained from table \ref{zredshiftvalues} (in the {$z$-redshift} space), while the red line represents equation \ref{Hubbletaylor} in terms of cosmographic parameters obtained from the last column of table \ref{tab:99slsParam} (in the {$y$-redshift} space).  {The vertical green line represents the convergence radius of the Taylor Series for the $y$-redshift case. \cite{Magana:2017nfs}}.}
    \label{fig:hdezvy}
\end{figure}

We conduct an additional test (using the MCMC bayesian analysis) in the $z$-redshift for the restricted sample. The values of the cosmographic parameters are reported in Table \ref{zredshiftvalues}. 
\begin{table}
    \centering
    \begin{tabular}{|c|c|}
    \hline
        Cosmographic parameter & Value obtained \\ \hline
        $q_0$ &  $-0.313\pm0.018$\\
        $j_0$ &  $1.83\pm0.045$\\
        $s_0$ &  $2.207\pm0.4656$\\
        $l_0$ &  $-0.5983\pm0.0544$ \\ \hline
    \end{tabular}
    \caption{Values for the cosmographic parameters up to 4th order in the $z$-redshift scenario, obtained with the set of 99 SLS data}
    \label{zredshiftvalues}
\end{table}
The aforementioned parametrization presents fitting 
 difficulties in zones where the redshift $z$ is away from the origin ($z=0$), in order to avoid this problem, we need a parameterization that can map the whole redshift spectrum into a closer region where the expansion is made (see \cite{Lizardo:2020wxw} for details). Here, we employ the so-called $y$-redshift parametrization, first proposed by Celine Cattoen et. al. in \cite{Catto_n_2007}, given as,
\begin{equation}
    y = \frac{z}{1+z},
\end{equation}
expressing the analogous equation to \ref{Hubbletaylor} in terms of the $y$-redshift, we obtain
\begin{eqnarray}
    H(y) & = &  H_0 \Big[ 1 + (1+q_0)y +\frac{1}{2}(2+2q_0-q_0^2 + j_0)y^2 \nonumber \\
     &+& \frac{1}{6}(6+6q_0-3q_0^2 + 3q_0^3 -4q_0j_0 + 3j_0 -s_0)y^3 \nonumber \\
     &+& \frac{1}{24}(24-4j_0^2+l_0+24q_0-12q_0^2+12q_0^3-15q_0^4
     \nonumber\\
     &+& j_0(12-16q_0+25q_0^2)-4s_0+7q_0s_0)y^4 \nonumber \\  &+&\mathcal{O}(y^5) \Big], \label{Hubbley}
\end{eqnarray}
which is the Hubble parameter in terms of the {$y$-redshift}, expanded as a taylor series around $y=0$. An  advantage of transitioning to the $y$-redshift space, is that it remains valid for higher values of the redshift. This is illustrated in Figure \ref{fig:hdezvy}, where we compare the Friedmann equations in terms of cosmographic parameters obtained through {$z$-redshift} and {$y$-redshift} parametrizations (through an MCMC analysis), providing a better fit to the data from cosmic chronometers (presented in \cite{Magana:2017nfs}) in the {$y$-redshift} space up to $z \sim 1.3$\footnote{This improvement comes with the cost of increasing the errors in the estimation of the cosmographic parameters in comparison with those obtained in $z$-redshift (see also~\cite{Busti:2015xqa} ).} . Therefore, we have selected this parameterization because SLS encompasses higher redshifts than those presented in cosmic chronometers measurements.

\section{DHOST Lagrangians}\label{appendix:b}
Considering DHOST with up to quadratic terms in second order derivatives of the scalar field, the general Lagrangian has the form
\begin{equation}
{\cal L}_{quad} \,=\,G_2(X) + G_3(X) \Box\phi+\sum_{i=1}^5{\cal L}_i+G(X) R +\mathcal{L}_m\,, \label{ESTlag}
\end{equation}
where $X=\phi_\mu\phi^\mu$, $R$ is 
the Ricci scalar of the metric $g_{\mu\nu}$, $\mathcal L_m$ is a Lagrangian for matter minimally coupled to the metric, and the Lagrangians $\mathcal L_i$ are defined as
\begin{align}
{\cal L}_1 [A_1] &  = A_1(X) \phi_{\mu \nu} \phi^{\mu \nu} \,,\label{A1} \\
{\cal L}_2 [A_2] & = A_2(X) (\Box \phi)^2 \,,
\label{A2} \\
{\cal L}_3 [A_3] & = A_3(X) (\Box \phi) \phi^{\mu} \phi_{\mu \nu} \phi^{\nu} \,,\label{A3} \\ 
{\cal L}_4 [A_4] & = A_4(X)  \phi^{\mu} \phi_{\mu \rho} \phi^{\rho \nu} \phi_{\nu} \,, 
\label{A4}
\\ 
{\cal L}_5 [A_5] & = A_5(X)  (\phi^{\mu} \phi_{\mu \nu} \phi^{\nu})^2\,.
\label{A5}
\end{align}
 The functions $G, G_2, G_3, A_i$ are arbitrary functions of $X$. This last point is assumed for simplicity; in principle these functions might also depend on $\phi$. 
In general, gravitational waves in this theory propagate at a speed different from that of electromagnetic waves. To avoid this, it is necessary to set $A_1 = 0$. Furthermore, the Ostrogradski ghost is avoided if $A_2= 0$ and~\cite{Crisostomi:2017pjs}
\begin{align}
A_4 &  = -\frac{1}{8G}(8A_3 G -48G_X^2 - 8 A_3 G_X X + A_3^2 X^2), \nonumber \\
  A_5 & =\frac{A_3}{2G}(4G_X+A_3 X).
\end{align}
Following~\cite{Crisostomi:2017pjs}, the free functions $G, G_2, G_3$ and $A_3$ are chosen as
\begin{equation}
 G = \frac{M_P^2}{2}+\frac{c_4}{\Lambda_3^6} X^2 \,,\quad  G_2 = c_2 X \,, \quad G_3 = \frac{c_3}{\Lambda_3^3}X \,,  \quad A_3 = - \frac{8 c_4}{\Lambda_3^6} - \frac{\beta}{\Lambda_3^6} \,.\label{functs}
\end{equation}
With these choices, the model under consideration reduces to the covariant galileon of beyond Horndeski when $\beta = 0$.

The equations of motion are obtained by varying the Lagrangian with respect to the metric and scalar field, and then imposing the Ansatz that the line element describes a flat Friedmann-Robertson-Walker spacetime,

\begin{align}
ds^2 = -dt^2 + a(t)^2(dr^2 + r^2 d\theta^2 + r^2 \sin^2\theta d\varphi^2),
\end{align}
and that the scalar field depends only on the time coordinate.  
As described in~\cite{Crisostomi:2017pjs}, this leads to three independent equations that contain up to fourth derivatives of the scalar field and second derivatives of the Hubble parameter. However; the spatial part of the field equations for the metric can be solved for $H'$, and after substituting the result in the remaining equations one finds two second order equations that only contain $\phi, \phi'$ and $H$.
Then we look for numerical solutions to the two independent equations,
 \begin{align}
 0&=\dot\phi^{12} \left\{8 c_{24} \left[\left(9\beta\beta_{12} - 32 c_4^2 \right) \ddot\phi+24 c_{34}\right]+3 \ddot\phi \left[9 \beta ^2\beta_{16}\beta_{20}  \ddot\phi^2+72 \beta  c_{34}  \beta_{16} \ddot\phi+128 c_{34}^2\right]\right\}
 \nonumber\\
 &+6 \dot\phi^8 \left[2 c_2 \left(\left(3 \beta ^2-8 c_4 \beta_{12}\right) \ddot\phi +16 c_{34} \right)+\ddot\phi \left(9 \beta  \beta_{16} \ddot\phi +32 c_{34}\right) \left(2 c_3-\beta  \ddot\phi\right)\right]-32 c_2 \ddot\phi\nonumber \\
 &+24 \dot\phi^4 \left[4 c_3 \ddot\phi (c_3-\beta \ddot \phi)+ 2 c_{23}-c_2 \left(5 \beta +8 c_4\right) \ddot\phi\right] \nonumber \\
 &+36 H^2 \dot\phi^2 C_1^2 \left\{3 \beta_{16} ( \beta_{20} \dot\phi^4-2)\ddot\phi+4 c_3 C_1\right\} \nonumber \\
 &-12 H \dot\phi C_1 \left\{4 c_2 C_1 (3\beta_{12} \dot\phi^4+2)+\left[8 c_3 C_1 (3 \beta_{12} \dot\phi^4-2)  + 3 \beta  \beta_{16} \dot\phi^4 \ddot\phi (3 \beta_{20} \dot\phi^4-6)\right]\ddot\phi \right\}, \\
 0&=3 C_1^{-1}\beta ^2 \dot{\phi }^6 D_{20} \ddot{\phi }^2+24 \beta  c_3 \dot{\phi }^6 \ddot{\phi }+4 c_2  \dot{\phi }^2 D_4-12  H \dot{\phi }^3 \left(\beta  D_{20} \ddot{\phi }+4 c_3 C_1\right)+12 C_1 H^2 D_{20},
 \end{align}
 where 
 \begin{align}
 C_1& =  2c_4 \dot\phi^4+1, \ \ D_{4} = 3 \beta _4 \dot{\phi }^4+2, \ \ D_{20} = 3 \beta _{20} \dot{\phi }^4+2, \nonumber \\
 \beta_{12} & =\beta+4 c_4, \ \ \
 \beta_{16}  = \beta + 16 c_4/3, \ \ 
 \beta_{20}  = \beta + 20 c_4/3,  \nonumber \\  c_{ij}&=c_i c_j,
 \end{align}
 and we introduced the dimensionless quantities $t\to M _P^{1/2}\Lambda_3^{-3/2}t, H\to M _P^{-1/2}\Lambda_3^{3/2}t, \phi\to M_p\phi$.         
\end{document}